\documentclass{mem}
\usepackage{natbib}\usepackage{txfonts}\usepackage{balance}
\usepackage{graphicx}
\usepackage[a4paper,breaklinks,dvipdfm]{hyperref}
\idline{75}{282}
\begin{document}
\def\teff{$T\rm_{eff }$}
\def\kms{$\mathrm {km s}^{-1}$}

\title{Thirty years of pulsar studies at ESO. \\The Italian Contribution.
}

   \subtitle{}

\author{
R. P. Mignani\inst{1,2}
          }

  \offprints{R. P. Mignani}

\institute{Mullard Space Science Laboratory, University College London, Holmbury St. Mary, Dorking, Surrey, RH5 6NT, UK
\and
Kepler Institute of Astronomy, University of Zielona G\'ora, Lubuska 2, 65-265, Zielona G\'ora, Poland 
\email{rm2@mssl.ucl.ac.uk}
}

\authorrunning{R. P. Mignani}

\titlerunning{Pulsar studies at ESO}

\abstract{In May 1982, when Italy joined ESO, only two isolated neutron stars (INSs) had been identified in the optical: the Crab and Vela pulsars. Thanks to the ESO telescopes and the perseverance of a few Italian astronomers, now about 30 INSs have been identified in the optical/IR, and a new important channel in their multi-wavelength studies  has been opened. In this contribution, I review the major steps in 30 years of INS studies at ESO, highlight the role of Italian astronomers, and introduce future perspectives with the E-ELT.

\keywords{Stars: pulsars}
}
\maketitle{}

 In 1982 about 330 isolated neutron stars (INSs) had been identified as radio pulsars. Two (Crab and Vela) had also been detected  in $\gamma$-rays  by the NASA's {\em SAS-2}  (1972-1973) and the ESA's {\em COS-B} (1975-1982) satellites and ten were detected in X-rays by the  {\em Einstein} Observatory.  Thus, pulsars had become targets for multi-wavelength astronomy. In the optical, Crab (V=16.5) and Vela (V=23.6) had also been detected as optical pulsars (Cocke et al.\ 1969; Wallace et al.\ 1977). 
 A third one (PSR\, B0540$-$69; V=22.5) had been discovered in the LMC (Middleditch \& Pennypacker 1985) but its counterpart was unidentified against its spatially unresolved SNR.  
 At the end of the 1990s, the quest for the understanding Geminga (Bignami \& Caraveo 1996) triggered new interest in optical observations of INSs.  Geminga, an unidentified $\gamma$-ray source discovered by SAS-2 and COS-B in the Galactic anti-centre (in the Gemini constellation, hence the name) had been suspected since its discovery  to be the third Galactic $\gamma$-ray pulsar, after Crab and Vela. However, due to the paucity of  $\gamma$- photons, no $\gamma$-ray pulsation had been detected at the time, and no radio pulsar had been found  in the $\gamma$-ray error box. This left to multi-wavelength observation campaigns the task of ascertain the nature of Geminga. The discovery of a candidate X-ray counterpart with {\em Einstein} triggered optical follow-ups  and a candidate  counterpart (G") was soon detected by the CFHT and  the ESO/3.6m (Bignami et al.\ 1987; 1988). Its faint flux (V=25.5) compared to the X-ray one implied an X-ray--to--optical flux ratio in excess of 1000, virtually suggesting that Geminga was indeed an INS, the fourth to be identified in the optical (not yet in radio, though). The 1990s saw new high-energy observing facilities, such as the {\em Compton Gamma-ray Observatory} (1990-2000) and {\em ROSAT} (1990-1999), taking over from {\em COS-B} and {\em Einstein}.  A total of seven pulsars were then detected in $\gamma$-rays and about 30 in X-rays. In the optical, the NTT became operational at the ESO La Silla Observatory in 1989.  Thanks to its unprecedented performances, the NTT was crucial to secure more pulsar identifications (Mignani et al.\ 2000). First of all, it secured the optical identification of Geminga  through the optical proper motion of its counterpart G" (Bignami et al.\ 1993).  Then, NTT observations yielded to the identification of the LMC pulsar PSR\, B0540$-$69,  the discovery of the counterpart (V=25) of the middle-aged pulsar PSR\, B0656+14, tentatively detected in previous 3.6 images, and of a candidate counterpart for the young pulsar PSR\, B1509$-$58 (Caraveo et al.\ 1992; 1994a; 1994b). Many more pulsars were also investigated with the NTT for the first time (Mignani et al.\ 2000). Riding the wave, the NTT  provided the first CCD spectrum of the Crab pulsar, the first tentative measurement of the secular decrease of its optical luminosity (Nasuti et al.\ 1996), predicted by Pacini \&Salvati (1983),  and a robust measurement of the Vela pulsar proper motion  (Nasuti et al.\ 1997),  which confirmed the SNR association. 
In the mid 1990s, most pulsar identifications had been obtained by ESO telescopes, mainly the NTT, and thanks to Italian astronomers, mainly at the IFCTR in Milan (now INAF-IASF). These results gave ESO a leading role in studies of southern pulsars, till then dominated by the AAT and CTIO,  and were seminal for follow-ups with the refurbished {\em HST} (Mignani 2010), also started by the Milan's group. Moreover, it spurred more and more interest in the Community, involving many more groups the world over, and consolidated the optical as an important branch for INS astronomy. \\
At the end of the 1990s, the VLT became the flagship of the ESO telescopes fleet and took over the NTT role. After the first light of the VLT first unit telescope (UT1, later Antu), pulsar observations were accepted as a test case for the telescope science verification, and the results of the first pulsar observations with the VLT were published in the Astronomy\&Astrophysics Special Edition (Mignani et al.\ 1999).  VLT observations, spurred by the competition with the {\em HST},  continued on the path paved by the NTT (Mignani 2009) and yielded to new pulsar identifications (e.g., Mignani et al.\ 2008), the first measurement of the optical spectrum and polarisation of the Vela pulsar (Mignani et al.\ 2007a,b),  the first detections of pulsars in the near-IR  (e.g., Mignani et al.\ 2012a), and, more recently, the first optical surveys of pulsars detected in $\gamma$-rays by {\em Fermi} (Mignani et al.\ 2011, 2012b).
ESO observations have also been crucial to study the properties of pulsars in binary systems, in particular  the so-called milli-second pulsars (MSPs),  a research led by groups at the Bologna University and Observatory and at the Cagliari Observatory. MSPs are billion year old pulsars which have radiated away most of their rotational energy during their life time and, after going through a phase of radio silence, have been spun up to milli-second periods through matter accretion from a binary companion and became active radio pulsars once again.  While the pulsars themselves are obviously too faint to be detected against their brighter companions, optical observations allow one to study the characteristics of the latter (usually late main sequence stars or white dwarfs) and study the system properties and evolutionary path. In particular, optical observations allow one to study the orbital variability of the companion star and determine the orbit inclination angle, the system masses from the radial velocity curves, and the chemical composition, temperature, luminosity, and age of the companion. All these parameters are important to trace back different steps in the accretion phase and the MSP spin up. Moreover, the study of the optical variability is crucial to investigate the effects of the companion irradiation from the MSP,  a phenomenon observed in the so-called Black Widow pulsars, and constrains the pulsar inclination angle with respect to the orbital one and the radius of the companion star (e.g. Pallanca et al.\ 2012). 
At the same time, VLT observations have been instrumental to determine the nature of some peculiar radio-silent INSs discovered at high energies. One case is that of the so-called Soft Gamma-ray Repeaters (SGRs). The prototype of this class (SGR\, 0529$-$66) was discovered on March 5, 1979 in the LMC as a source of recurrent $\gamma$-ray bursts.  While the burst recurrence pointed at a source different from the already known GRBs, the source position, coincident with the $\approx$ 2000 year old N49 SNR and the detection of 8s pulsations suggested that it was a spinning INS, albeit with an unusually long period for its young age. More or less contemporarily, came the discovery of a puzzling class of presumably isolated X-ray pulsars also associated with young SNRs, dubbed {\em Anomalous} (a.k.a. AXPs) because of their long spinning periods (also in the few s range) and X-ray luminosity larger than the neutron star rotational energy reservoir, as estimated from the period and period derivative assuming a magnetic dipolar spin-down. This was shown to be the case also for the SGRs, once their steady and pulsating X-ray counterparts were discovered and monitored (Mereghetti\ 2008). Thus, it was clear that both classes had to be powered  by a source different than the neutron star rotation. Interestingly enough, their inferred magnetic fields were in the $10^{14}$--$10^{15}$ G range, a factor of 10-100 stronger than in radio pulsars. Both SGRs and AXPs, were proposed to be {\em magnetars}, hyper-magnetised INSs powered by their magnetic field. The magnetar interpretation, however, has been for a long time debated and models based on accretion from either low-mass stars or debris disk formed out of the supernova explosions were invoked, at least to explain the steady X-ray emission of both SGRs and AXPs and the lack of indirect observational evidence from the presence of massive companion stars (e.g. X-ray eclipses, Doppler shifts of the X-ray pulsation, etc.). Of course, optical/IR observations have been crucial to discriminate between different models. Unfortunately, the location of most magnetar candidates in the Galactic plane, i.e. in regions of high crowding and extinction, hampered for a long time the search for their optical/IR counterparts, which had to rely on accurate X-ray positions and deep high spatial resolution IR imaging.  These became possible thanks to {\em Chandra} and Adaptive Optics IR imagers, such as NACO at the VLT. Moreover, the rapid response of the VLT was key to quickly activate Target of Opportunity observations triggered by the detection of X or $\gamma$-ray bursts from these sources by  {\em Swift} and {\em RXTE} and spot their bursting IR counterparts. In this way, counterparts to several magnetar candidates were discovered (see Mignani 2009 and references therein), mainly thanks to the work of colleagues at the Rome Astronomical Observatory, and was possible to set tight constrains on accretion from a companion star or a debris disk and support the magnetar model.   Moreover, indirect evidence has been found that the IR emission, like the X-ray one, is powered by the magnetic field (Mignani et al.\ 2007c). 
Another group of X-ray bright radio-silent INSs were discovered in the 1990s in  the{\em ROSAT} all-sky survey  (Turolla\, 2009). These sources, not in SNRs, are characterised by soft X-ray thermal spectra. Initially, they were thought to be old no-longer active radio pulsars, whose X-ray emission was ascribed to the re-heating of the neutron star surface by ISM accretion. Optical observations were crucial to ascertain that these sources were INSs, while the detection of X-ray pulsations only came after the launch of {\em XMM-Newton} in 1999. Indeed, deep NTT observations, lead also by  groups at the Padua's University and Observatory, allowed one to determine that their X-ray--to--optical flux ratio was in excess of 1000 and that they were INSs, a.k.a X-ray Dim INSs (XDINSs) for their dimness in the {\em ROSAT} data. Moreover, the identification of the first XDINS optical counterpart (RX\, J1856.5$-$3754) allowed one to infer, through its optical proper motion,  the space velocity relative to the ISM. This was too high for efficient accretion and proved that the XDINSs' thermal emission was purely due to the original cooling of the neutron star surface. XDINSs' observations with the VLT also lead to the identification of the counterparts for a few of them (Zane et al.\, 2008; Mignani et al.\ 2009). \\
Thus, one can certainly say that INS optical astronomy grew thanks to the enduring work of Italian astronomers, including a few expatriates. Indeed, about half of the 30 INS detected in the optical/IR have been identified/studied at ESO by Italian astronomers. While the NTT represented the beginning and the VLT the glorious continuation, the E-ELT represents a new era in INS optical astronomy.  Its unprecedented light-collecting area will yield many more identifications, reducing the gap with X-ray and $\gamma$-rays, and produce the large photon fluxes required for polarimetry, spectroscopy, and timing observations, so far limited to a few objects only but needed to make progress in our understanding of the INS emission physics. In particular, E-ELT spectroscopy (METIS, MICADO, OPTIMOS) will allow one to study the INS optical spectra, separate thermal/non-thermal, components, measure the bulk of the neutron star surface temperature, detect cyclotron spectral features in the magnetosphere or absorption features in the atmosphere, and search for debris disks. E-ELT polarimetry (EPICS) will allow one to measure the atmosphere/magnetosphere magnetic field, build pulsar-wind nebulae polarisation maps, determine the average polarisation orientation with respect to the neutron star spin axis and proper motion. E-ELT astrometry (MICADO) will allow one to measure proper motion of radio-silent INSs up to the LMC and parallaxes at $ >$1 kpc. Finally, timing observations with the E-ELT will be crucial to study very fast variability phenomena, such as giant pulses, pulse drifts, intermittent pulses, bursts, in synergy with facilities like the SKA and {\em LOFT}. Again, the Italian community is riding the way.  A new generation of high-time resolution instruments down to the pico-s, based on quantum detector  technology, is being developed at the Padua's University and Observatory, and a prototype (IQuEye) has been successfully tested at the NTT. A VLT version (VQuEye) is now under study (PI C. Barbieri) and an instrument concept proposal has been submitted to ESO for an E-ELT prototype (EQuEye), which will open a new time domain to optical astronomy. 
 
\begin{acknowledgements}
The author thanks OPTICON  for financial support.
\end{acknowledgements}

\bibliographystyle{aa}

\end{document}